# Machine Learning: Challenges, Limitations, and Compatibility for Audio Restoration Processes

Owen Casey[#1], Rushit Dave[#2], Naeem Seliya[#3], Evelyn R Sowells Boone[*4]

[#]*Department of Computer Science, University of Wisconsin-Eau Claire*
*105 Garfield Ave, Eau Claire, WI 54702, United States*

[*]*Department of Computer Systems Technology, North Carolina A&T State University*
*Greensboro, NC 27411, United States*

[1]`caseyo17411@uwec.edu`

*Abstract*— In this paper, machines learning networks are explored for their use in restoring degraded and compressed speech audio. The project intent is to build a new trained model from voice data to learn features of compression artifacting (distortion introduced by data loss from lossy compression) and resolution loss with an existing algorithm presented in 'SEGAN: Speech Enhancement Generative Adversarial Network'. The resulting generator from the model was then to be used to restore degraded speech audio. This paper details an examination of the subsequent compatibility and operational issues presented by working with deprecated code, which obstructed the trained model from successfully being developed. This paper further serves as an examination of the challenges, limitations, and compatibility in the current state of machine learning.

*Keywords*— Machine Learning, Audio Processing, Speech Enhancement, Deep Learning, Generative Adversarial Networks, GPU, CUDA, Tensorflow, PyTorch, Keras

## I. Introduction

With rare exception, every electronic system that sends and/or stores files utilizes some form of data compression, to save on storage and bandwidth.

Compression has two fundamental types, lossless and lossy. Lossless compression leverages algorithms and efficient designs to reduce the data required to represent the file, without losing any of the original data. A lossless file can be completely restored to its original. Lossy compression, however, removes data but targets data that is negligible or minimally disruptive [1]. MPEG-1 Layer-3 or Mp3 is a widely adopted form of lossy compression. It reduces files massively but at the cost of losing audio data. The encoder, among other optimizations, leverages humans' perception of sound (psychoacoustics) and discards parts of the signal such that their removal is imperceivable to the human ear [2]. An example is low passing (filtering out / removing above a given frequency) audio frequencies past 20kHz, since anything above is out of the human hearing range [3].

When done effectively, lossy compression can drastically reduce the size of a file, with minimal to no discernible loss in quality, but lossy compression becomes problematic with repeated resampling and compression or a single compression that is too aggressive in data removal. In these cases, compression artifacts and resolution loss become apparent. A perhaps relatable example might be popular videos that have been distributed, re-edited and re-uploaded many times [4]. The repeated compressions iteratively reduce the quality of the sound and video.

Audio Processing with machine learning offers solutions to problems that programmatic solutions could never come close to [5], 6]. Deep networks ability to dynamically learn features over large training sets allow a wider range of conditions and characteristics to be addressed [7]. This dynamic learning makes deep networks flexible and variable for more use cases [8], [9].

This research is to build a new model using an existing machine learning project, SEGAN: Speech Enhancement Generative Adversarial Network [10]. While the original objective of SEGAN was to remove background noise from voice audio, the intent was to use the network to learn the features of compression artifacting and resolution loss.

SEGAN is a Generative Adversarial Network, meaning the network involves two networks, a generator and a discriminator. The generator is a network that aims to generate an intended output and the discriminator is a network that aims to classify a given input as real (genuine input data) or fake (a piece of data generated by the generator). Both models train together and compete. The generator to fool the discriminator, and the discriminator to catch the fakes from the generator. As these models train, the discriminator improves its classification and the generator gets better at making convincing output.

The model would have been trained with a dataset created with source audio from the Mozilla Common Voice data set [11], then processed into two mirrored datasets, one of raw, clean voice audio and the other of that same data compressed with lossy compression down to 32kbps. parameters. The intent was that the model would learn the features of compression artifacting and thus be able to output clean, uncompressed audio from degraded audio using the trained generator. More specifically, the intent would be to learn the features generally, such that it could be generally applied and improve resolution over data that was not explicitly degraded by mp3 or lossy compression.

## II. METHODS

The intended trained model was built from a constructed voice dataset, learning the features of compression artifacting and quality reduction, structured from SEGAN [10] as the machine learning network.

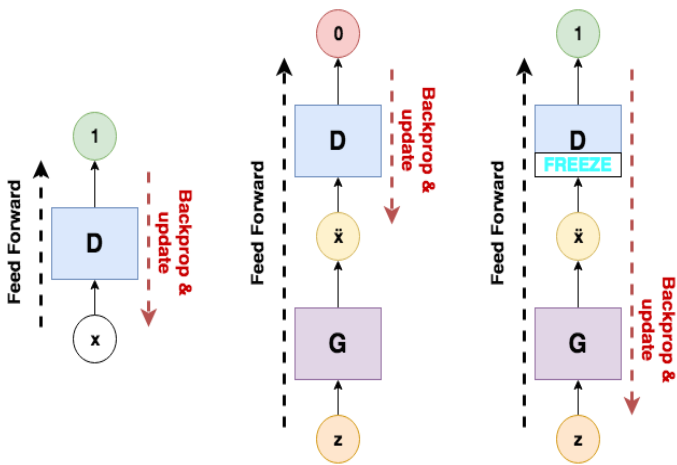

Figure 1: *The discriminator (D) and generator (G) network and their training function. Their weights and biases of the networks are adjusted with backpropagation. Initially, D is trained on real input data. G is then fed input and D classifies the output from G. If D's classification is incorrect D's parameters are updated. If D is correct, then G is adjusted [10].*

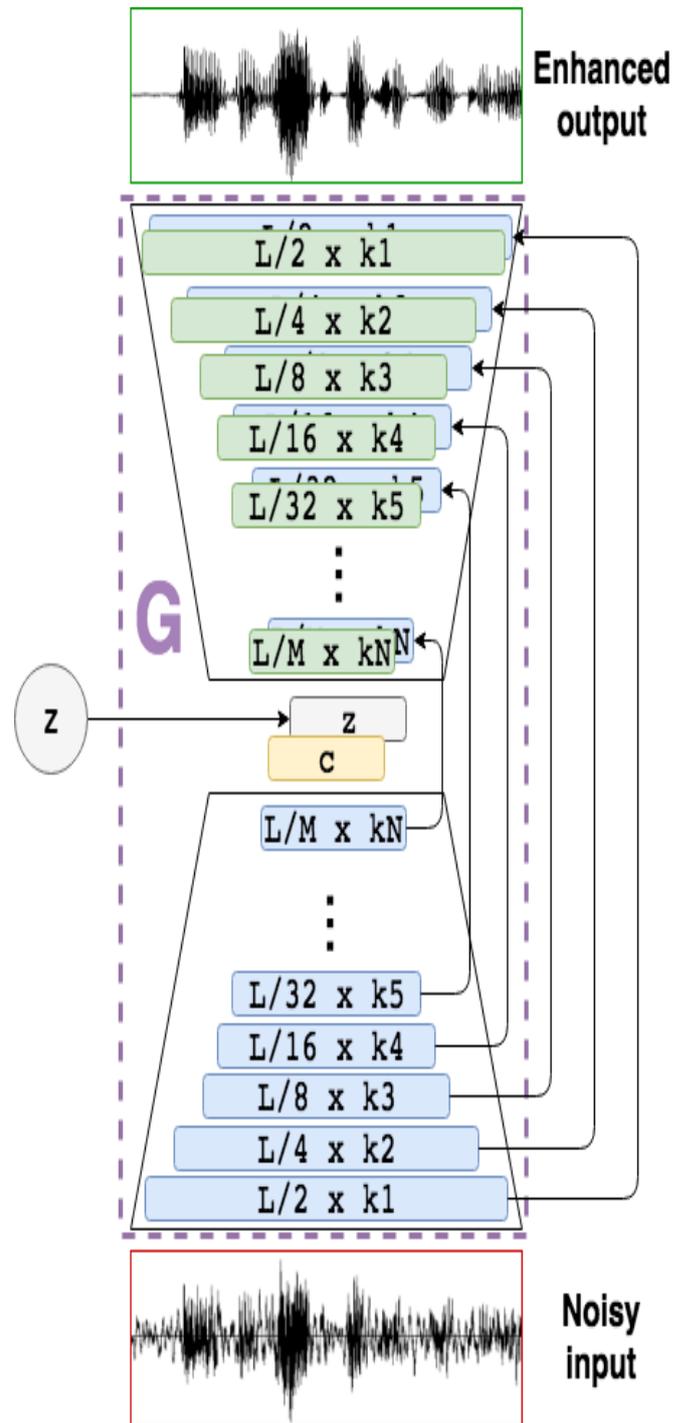

Figure 2: *An overview of the encoding/decoding architecture of the generator network G. The generator is composed of two sets of kN convolutional layers, which projects the input into a more condensed format. The resulting vector c is*

*concatenated onto the vector z, and then processed through kN decoding convolutional layers [10].*

The source dataset is a 30-hour subsection of the Mozilla Open Voice dataset [11]. The source set consists of short, mostly English phrases, between 2-6 seconds from a wide variety of speakers. The full dataset is composed of two mirrored sets, clean and compressed. The compressed dataset is generated from the clean dataset using the compression plug-in VST (Virtual Studio Technology) "Lossy" [12] which offers a variety of functions to control lossy bitrate compression. The dataset is batch processed using "Wavosaur", a DAW (Digital Audio Workstation) with VST batch process functionality, allowing processing to be applied to folders of audio at a time. Both sets of audios were resampled to a 16kHz sample rate and mono to be compliant with the input requirements of SEGAN. This trained model was not built due to restrictions of deprecated code and operational issues.

Limitations and conflicts of SEGAN and machine learning for audio processing tasks are discussed in the next section.

### III. LIMITATIONS, CONFLICTS, AND RESULTS

*A. Options for Executing Machine Learning Code*

- ***Local System and GPU:*** It is effective to run code directly on a local machine using local hardware and GPU. It also requires sufficient storage, ram, CPU power, cooling, and any other potential bottlenecks, like file read and write speed. A GPU is not technically required to process, but it drastically improves performance.

- ***Hosted System and GPU:*** Paying a third party to use their hardware and deploying environments in a wrapped virtual environment [13]. Remote systems configurations range from simple deployment and terminal access to dynamically deployed ipython notebooks integrated with dynamically allocated resources as seen in Google Collab [14].

- ***Simplified models on CPU locally:*** Simplifying the model to a lower accuracy and / or resolution at the benefit of less processing demand. Simple enough models and / or powerful enough hardware can be run in browsers or integrated into applications [15].

- ***Server Backend:*** Taking input from the user, managing the processing demands and configuration requirements on the server end, and then returning the output to the user.

*B. Fundamental Architectures, Platforms and Libraries*

GPUs are often used in machine learning to expedite what would take a CPU many times longer. CPUs are generalized and can compute any task, but this makes them slower than specific hardware. GPUs have specialized architecture that allows specific tasks to compute quickly and efficiently. Leveraging this architecture, neural network primitives like neurons can be abstracted and processed much faster on a GPU [16].

The most significant API is Cuda, Nvidia's system for connecting Nvidia GPUs for general computing tasks. OpenCL is a potential alternative, but lacks the support, adoption and performance of Cuda [17]. Notable libraries include Tensorflow, [18] Google's codebase for machine learning structures, PyTorch, [19] a python-based library for machine learning, and Keras, [20] another python library with dynamic backend support.

These systems are constantly changing and upgrading, as well as their dependencies and lateral (frequently used in conjunction) dependencies. This makes compatibility difficult. The older the code, the more dramatic the compatibility and support issues are.

*C. Compatibility Issues*

- ***Dependencies:*** For the code to function correctly, often the installed dependencies need to be the exact version the code was written with. When the specific dependency version is inaccessible, conflicts with other dependencies, or more often, simply are not specified; this can cause errors and require reimplementation to correct.

- *Inputs and settings:* In order for the code to run and not fail during runtime, sometimes settings need to be adjusted to fit your configuration or given input and will simply break if things are not set up correctly. If error reporting is not clear, or well designed, it makes diagnosing the problem difficult and requires a more intimate understanding of the source code to remedy.

- *Driver, Operating System and Environment Compatibility:* Compatibility between the hardware and the drivers, the code to interface the hardware and software together, are required and can introduce an additional layer of compatibility conflicts. Additionally, drivers are not always written for all types of operating systems. Sometimes a specific Operating system or environment to run the code is needed.

D. Novel field and Support Issues with Open-Source Code

- *Novel Field:* Only recently has processing become sufficiently powerful, and our data collection so effective and massive, that machine learning has only really been viable for the last few years. As such, the field is not as established as other subjects, which makes troubleshooting more difficult as there are less resources and information when compared to more established fields. The landscape and tools are changing rapidly. Resources fade quickly in relevance as new methods and systems supplant each other. Furthermore, with a newer field, less pieces of laterally helpful information (information that does not directly give the solution to your problem but can be used to solve the issue) are available, unlike more established practices with a greater set of related support and documentation.

- *Code documentation varies:* Even with disciplined developers, documentation when writing code is often lower priority. Therefore, documentation for open-source code can vary from professional documentation, FAQs, support pages to zero comments in the code.

- *Less robust code and less informative errors:* Public code is often more unstable. Errors often do not give useful information. For example, an error may seem like a syntax error, but is instead an error from improperly formatted input.

E. Limitations / Processing demands

- *Resolution:* The greater the resolution, the more required outputs and inputs, thus the greater the processing demand. As such, often resolution is compromised to make processing demand less. With respect to restoration, resolution is key, its bottleneck in audio models further complicates the task of audio restoration to an acceptable degree.

- *Model Complexity:* Number of parameters and layers in a model will increase the complexity for processing required data to train or run it. However, often the more layers are effective for the better result.

- *Large Data / Storage Requirements:* To build an effective model, large amounts of data are required. Collecting, and more importantly, filtering and formatting this data can be a difficult task. Storage demand to hold this data also adds to the problem.

- *Processing time:* Even when using a pretrained model and simply executing it often requires a significant amount of time to see the results. Adjusting quickly is difficult and especially so when compared to software processes that can complete many times per second.

F. Minimizing Processing

- *Simplify the model:* A model can be reduced in size by making a compromise of accuracy / clarity, and resolution [21].

- Resnet [22] demonstrated that processing can be reduced by using skip connections to jump layers and simplify the model while still maintaining equivalent or nearly equivalent results. SEGAN leverages skip connections within its generator architecture.

G. *Potential Solutions*

- ***Virtualization layers:*** Running in a virtualized environment like Docker [23] minimizes potential conflicts between the existing system and what the code requires to run. This can ensure a more consistent experience and maximize potential compatibility. Virtualization can be less efficient however, requiring more processing strain for the additional layer of abstraction. This can ensure the stability of the system and allow the deployment to be more system and hardware agnostic [24]. On the host machine, with driver connection to the virtual layer local hardware can be utilized while dynamically juggling a variety of virtual environments and configurations. This also makes managing multiple different configurations on a single machine convenient and easy to switch between [23].

- ***Well supported consumer facing application systems:*** With the right deployment, machine learning can be implemented directly into wrapped programs. This requires the compatibility burden to be put onto the developers, integrating and modifying the dependent systems to work in their ecosystem or sometimes implementing from scratch.

- ***Simplified models built into systems or platforms:*** Models with sufficiently small processing demands or leveraging smart systems to reduce demand for compromised or nearly representative equivalent accuracy can be built directly into conventional systems like web app environments given the right prerequisite dependency layers [25].

- ***Server Backend:*** Relaying user input from a web or application endpoint, processing server side, and returning the processed output.

- ***Increase in standard compute power:*** Nowadays the demanding nature of machine learning makes it non-viable for many real-time and consumer-grade applications. For many common applications like high-quality real-time video streaming or complex 3D live rendered game environments it can be used. If the standard of compute power for all devices increases, once our systems can handle a billion parameters with ease, these types of tasks can be more easily deployed.

| Execution | Limitation | Potential Solutions |
|---|---|---|
| Local GPU | Resolution | Virtualization Layers |
| Simple Local CPU | Complexity | Wrapped Applications |
| Hosted Computing | Data Size Requirements | Simple Local CPU |
| Server Backend | Processing Demands | Server Backend |

Table 1: *Execution, limitations, and potential solutions*

IV. CONCLUSIONS

Findings from SEGAN and other similar projects show the viability of machine learning for audio processing tasks. The flexibility of feature extraction could allow for models to dynamically learn features of audio, and inverse or directly apply these features. Other potential implementations are nearly endless.

Resolution is a significant hurdle in using machine learning for audio restoration / upscaling tasks. The results are drastically more effective than any equivalent programmatic solution; however, resolution directly increases model processing resource demand. The nature of compromising resolution for processing logistics makes these tasks only viable with long processing times and significant compute power. If compute power increases to a degree where these demands become trivial, neural networks will become viable as consumer grade systems that can be deployed and run with little to no configuration work, like modern web apps. Deprecated code and compatibility issues plague the usability of machine learning projects. Recent and supported projects require some work and knowledge to run, so depreciated projects complicate usability significantly.

REFERENCES

[1] M. Parahar, "Difference Between Lossy Compression and Lossless Compression." Tutorials Point. 08-Jan-2020. [Online]. Available: https://www.tutorialspoint.com/difference-between-lossy-compression-and-lossless-compression.

[2] K. Brandenburg, "MP3 and AAC Explained," 1999.

[3] D. Purves, Neuroscience, 2nd ed. Sunderland, MA: Sinauer Associates, 2001.

[4] A. Robinson, "Jpeg Artifacts vs Image Noise," 03-Apr-2011. [Online]. Available: http://blog.topazlabs.com/jpeg-artifacts-vs-image-noise/.

[5] Siddiqui N., Pryor L., Dave R. (2021) User Authentication Schemes Using Machine Learning Methods—A Review. In: Kumar S., Purohit S.D., Hiranwal S., Prasad M. (eds) Proceedings of International Conference on Communication and Computational Technologies. Algorithms for Intelligent Systems. Springer, Singapore. https://doi.org/10.1007/978-981-16-3246-4_54.

[6] Strecker S., Van Haaften W., Dave R. (2021) An Analysis of IoT Cyber Security Driven by Machine Learning. In: Kumar S., Purohit S.D., Hiranwal S., Prasad M. (eds) Proceedings of International Conference on Communication and Computational Technologies. Algorithms for Intelligent Systems. Springer, Singapore. https://doi.org/10.1007/978-981-16-3246-4_55.

[7] Menter Z., Tee W.Z., Dave R. (2021) Application of Machine Learning-Based Pattern Recognition in IoT Devices: Review. In: Kumar S., Purohit S.D., Hiranwal S., Prasad M. (eds) Proceedings of International Conference on Communication and Computational Technologies. Algorithms for Intelligent Systems. Springer, Singapore. https://doi.org/10.1007/978-981-16-3246-4_52.

[8] D. Gunn, Z.Liu, R. Dave, X. Yuan and K. Roy , " Touch-Based Active Cloud Authentication Using Traditional Machine Learning and LSTM on a Distributed Tensorflow Framework, " International Journal of Computational Intelligence and Applications(IJCIA),2019

[9] J. Shelton et al., "Palm Print Authentication on a Cloud Platform," 2018 International Conference on Advances in Big Data, Computing and Communication Systems (icABCD), Durban, 2018, pp. 1-6-IEEE.

[10] J. Mason, R. Dave, P. Chatterjee, I. Graham-Allen, A. Esterline, K. Roy, An Investigation of Biometric Authentication in the Healthcare Environment, Array, Volume 8,2020,100042, ISSN 2590-0056, https://doi.org/10.1016/j.array.2020.100042

[11] S. Pascual, A. Bonafonte, and J. Serrà, "SEGAN: Speech Enhancement Generative Adversarial Network," Interspeech 2017, 2017.

[12] "Common Voice Dataset," [Online]. Available: https://commonvoice.mozilla.org/en.

[13] "Lossy," Goodhertz, Inc." https://goodhertz.co/lossy.

[14] "Cloud GPUs (Graphics Processing Units)," Google, [Online]. Available: https://cloud.google.com/gpu.

[15] "Colaboratory FAQ," Google, [Online]. Available: https://research.google.com/colaboratory/faq.htm.

[16] X. Sun et al., "Training Simplification and Model Simplification for Deep Learning: A Minimal Effort Back Propagation Method," IEEE Transactions on Knowledge and Data Engineering, vol. 32, Feb. 2020.

[17] N. Lopes and B. Ribeiro, "GPUMLib: An Efficient Open-Source GPU Machine Learning Library," International Journal of Computer Information Systems and Industrial Management Applications, vol. 3, Jan. 2010.

[18] N. Dimolarov, "On the state of Deep Learning outside of CUDA's walled garden," 04-Jun-2019. [Online]. Available: https://towardsdatascience.com/on-the-state-of-deep-learning-outside-of-cudas-walled-garden-d88c8bbb4342.

[19] Tensor Flow, "09-Nov 2015. https://www.tensorflow.org/.

[20] "PyTorch," Sep-2016. [Online]. Available: https://pytorch.org/.

[21] "Keras," 27-Mar-2015. [Online]. Available: https://keras.io/.

[22] C. Bucila, R. Caruana, and A. Niculescu-Mizil, "Model compression," 2006.

[23] K. He, X. Zhang, S. Ren, and J. Sun, "Deep Residual Learning for Image Recognition." 2015.

[24] "NVIDIA Container Toolkit," 2015. [Online]. Available: https://github.com/NVIDIA/nvidia-docker.

[25] C. Tozzi, "Docker vs. Virtual Machines: Understanding the Performance Differences," 30-Nov-2016. [Online]. Available: https://www.channelfutures.com/technologies/docker-vs-virtual-machines-understanding-the-performance-differences.

[26] C. Thomas, "Deep learning based super resolution, without using a GAN," 24-Feb-2019. [Online]. Available: https://towardsdatascience.com/deep-learning-based-super-resolution-without-using-a-gan-11c9bb5b6cd5.